\documentclass[10pt,twocolumn,superscriptaddress,english,prl,aps,reprint,showpacs]{revtex4-1}
\usepackage{amsmath}
\usepackage{graphicx}
\usepackage{epstopdf}
\usepackage{caption}
\usepackage{subcaption}

\usepackage{subfig}

\usepackage{babel}

\begin{document}

\title{Exact results for deposition of binary mixtures of superdisks on the plane}

\author{N.M. \v{S}vraki\'{c}}
\affiliation{Institute of Physics Belgrade, University of Belgrade, Belgrade, Serbia}

\author{B. N. Aleksi\'{c}}
\affiliation{Texas A\&M University at Qatar, Doha, Qatar}

\affiliation{Institute of Physics Belgrade, University of Belgrade, Belgrade, Serbia}

\author{M. Beli\'{c}}
\affiliation{Texas A\&M University at Qatar, Doha, Qatar}

\begin{abstract}
We investigate the deposition of binary mixtures of oriented superdisks on a plane. 
Superdisks are chosen as objects bounded by $|x|^{2p}+|y|^{2p}=1$, where parameter $p$ controls their size and shape.  For single-type superdisks, the maximum packing and jamming densities are known to be nonanalytic at $p=0.5$. For binary mixtures, we discover that nonanalyticities form a locus of points separating "phase diagram" of shape combinations into regions with different excluded-area constructions. An analytical expression for this phase boundary and exact constructions of the excluded-areas are presented.

\end{abstract}


\maketitle

Covering of planar surface with differently shaped, non-overlapping geometrical objects is the subject 
of enduring interest to mathematicians, physicists, and engineers for both its theoretical and experimental importance. Theoretical studies have generally focused on topics such as maximum packing density \cite{jiaoStillingerTorquato}, stacking and clustering \cite{exactGen, Springer}, space filling \cite{spheres, TorquatoStillinger, Conway}, and other collective arrangements influenced by geometric features of basic building units. Almost all studies of this kind have considered convex units, usually circles or spheres.

On the experimental side, it is now possible to synthesize and manufacture particles (objects), ranging in size from sub-micrometer to nanometer scale \cite{synthesis}, with a wide assortment of well defined shapes and morphologies \cite{Sevonkaev, Goia}. Particles prepared in this way are often deposited on variously modified surfaces to improve and/or achieve new functionalities in applications in metallurgy, biomedicine \cite{Matijevic}, optoelectronics \cite{Andreescu} and other emerging fields of high technology \cite{Somorjai}. For instance, in 3D printing, the creation of three dimensional structure is achieved layer by layer, and the ability to control morphology, voids and other features of each individual layer is of great importance \cite{3Dprint, 3Dprint1}.  Similarly, in inkjet printing technology \cite{Goia}, nanosize silver particles are deposited on a surface to form a film which is subsequently sintered to produce electrically conducting structure with desired properties. The same experiments \cite{Goia} also revealed that the conductivity of the final product is improved if the pre-sintered film is formed using the \textit{mixture} of particles of two different sizes, instead of the single-size units. In parallel with these developments, important theoretical effort was devoted to understand and model key physical processes involved in particle formation, their nucleation and growth, size and shape selection, aggregation, etc.\cite{Gorshkov} 
 
Larger scale properties of the deposited film, e.g., coverage, jamming limit, or late stage kinetics of deposition, and their dependence on geometric features of constituent particles, are equally as important. For the purpose of such studies, two-dimensional Lam\'{e} objects (superdisks), defined as the set of points in the plane bounded by the curves $|x|^{2p}+|y|^{2p}=1$, with $p\in\left(0,\infty\right)$, that include large family of shapes, from concave ($p< 0.5$) to convex ($p> 0.5$), are particularly suited. Gromenko and Privman\cite{PRE79_042103gromenkoPrivman2009} have examined Random Sequential Adsorption (RSA) model \cite{evans} for deposition of superdisks on planar surface and have shown that jamming limit (i.e. the instance of coverage beyond which no further additions are possible) exhibits nonanalytic behavior when object's shape changes convexity, at $p=0.5$; additionally, it was shown \cite{Aleksic} that consequent nonanalyticities are observed in the late-stage kinetics of deposition at the same value of \textit{deformation parameter} $p$. An important recent work on optimal packing of superdisks \cite{jiaoStillingerTorquato}, for all values of $p$, reported similar results for nonanalyticities of densest packings. In these studies, the superdisks were all of the same type (shape), with a fixed value of deformation parameter $p$, for each individual instance of packing or coverage.
 
However, as hinted above, in some applications it is advantageous to use the \textit{mixture} of shapes (or sizes) for deposition. Also, deposited mixtures may sometimes exhibit novel behaviors, like spontaneous ordering, or size segregation, as reported in experiments with nanoscopic gold particles \cite{Nature}, which makes such deposition an interesting object of study in its own right. Unfortunately, for depositions of mixtures, what limited theoretical work there is is confined to the studies of objects of the same shape, such as binary mixture of circles with different radii \cite{Bouvard}, or mixture of simple line segments of different lengths \cite{svrakic, Cadilhe}. Here we report analytical and numerical results for Random Sequential Adsorption (RSA) model of deposition, by using binary mixture of different oriented Lam\'{e} objects to cover regular planar surface. To the best of our knowledge, this is the first study of this kind.

The advantage of using Lam\'{e} superdisks to study deposition of mixtures is in that many properties of interest can be obtained in analytic form and analyzed in considerable detail. Also, by controlling the value of single deformation parameter, $p$, the family of shapes changes from concave to convex (at $p=0.5$), and includes astroids ($p=1/3$), diamonds ($p=1/2$), circles ($p=1$), squares ($p=\infty$) and the complete range of intermediate forms, as illustrated in Fig. 1. For the chosen value of $p$, the surface area of the corresponding superdisk, $A(p)$, is given by 
\begin{equation}
A(p)=4\frac{\Gamma^{2}(1+\frac{1}{2p})}{\Gamma(1+\frac{1}{p})}\label{A_p}
\end{equation}
where $\Gamma$ is the Gamma function. Clearly, $A(p)$ is continuously increasing, analytic function with values $A(0)=0$, $A(0.5)=2$, $A(1)=\pi$, $A(\infty)=4$, etc.
 
\begin{figure}
\begin{minipage}{2.05cm}
\includegraphics[width=2.5cm,height=2.5cm]{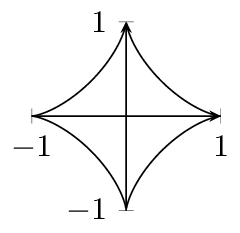}
\end{minipage}
\begin{minipage}{2.05cm}
\includegraphics[width=2.5cm,height=2.5cm]{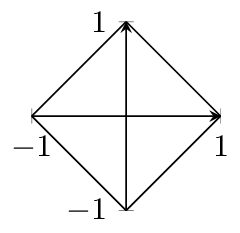}
\end{minipage}
\begin{minipage}{2.05cm}
\includegraphics[width=2.5cm,height=2.5cm]{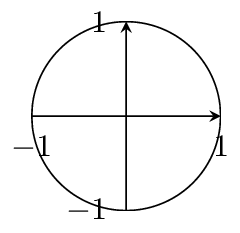}
\end{minipage}
\begin{minipage}{2.05cm}
\includegraphics[width=2.5cm,height=2.5cm]{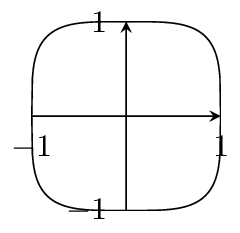}
\end{minipage}
\caption{Lam\'{e} objects (superdisks) with $p=0.3$, $p=0.5$, $p=1$, and $p=2$. Note the change in convexity at $p=0.5$}
\end{figure}

Deposition is realized following the usual rules of RSA model. In this model particles (objects) are sequentially deposited on the randomly chosen site on the substrate. When deposited, objects are irreversibly and permanently attached to that site. If the randomly chosen site for deposition is already occupied, or the object, if deposited, would overlap with any of its neighbors, the deposition is rejected, the object is discarded, and the deposition is next attempted at a different randomly chosen site. Note that, in this process, object-object and object-substrate interactions are modeled solely by geometrical and other features included in the deposition procedure.  With deposition of binary mixtures (two types of superdisks), in the first step, when the site for deposition is randomly selected, one must also decide which one of the two objects will be placed. In our numerical simulations, we choose one or the other with equal probabilities, but this can be altered at will.

The non-overlapping restriction implies that, in the neighborhood surrounding any deposited superdisk, there exists a region (the exclusion region) within which no center of another superdisk can be placed, i.e., depositions inside the exclusion region are not allowed. The size and shape of the exclusion region (ER) depends on the size and shape of the already deposited superdisk and on the size and shape of the superdisk we are attempting to deposit. Finding the exact shape and area of the ER is thus the central issue in the studies of packing and deposition \cite{jiaoStillingerTorquato,PRE79_042103gromenkoPrivman2009,Aleksic}.

In the "pure" case (i.e. deposition of single-type superdisks), the ER are well known \cite{PRE79_042103gromenkoPrivman2009,Aleksic}. For convex superdisks ($p>0.5$), the ER is constructed by simple rescaling of the original superdisk by a factor of 2, i.e., the ER of the superdisk $|x|^{2p}+|y|^{2p}=1$ is the set of points bounded by $|x|^{2p}+|y|^{2p}=2^{2p}$, as illustrated in Fig. 2(a). Within this region no center of another superdisk can be placed. The area of this ER is $\mathcal {A}_{cx}=4A(p)$ where $A(p)$ is given by eq. (1), and subscript indicates "convex" case (we call this "the convex construction"). For concave superdisks ($p<0.5$), ER is geometrically constructed by unit translation of the original superdisk in $\pm x$ and $\pm y$ directions. The set of points encompassed within the external envelope of these four superdisks is the ER of the original object, as illustrated in Fig. 2(b). The envelope of this ER can be parametrically written as $x(t)=\pm(1+t^{2p})^{-\frac{1}{2p}}\pm 1$, $y(t)=\pm t\times (1+t^{2p})^{-\frac{1}{2p}}$, with similar (four) expressions in which $x(t)$ and $y(t)$ are interchanged, and $t\in (0,\infty)$. The area of this ER is easily derived and is given by $\mathcal {A}_{cv}=4+2A(p)$ where $A(p)$ is given by eq. (1), and subscript indicates "concave" case (we call this a "concave construction"). Clearly, as a function of $p$, the \textit{area} of the ER, $\mathcal{A}$, is continuous function for all values of $p$, but has discontinuous first derivative at $p=0.5$. This nonanalyticity is then reflected in the nonanalyticities of the maximum packing density \cite{jiaoStillingerTorquato} as well as jamming limit \cite{PRE79_042103gromenkoPrivman2009} and the late-stage kinetics of deposition \cite{Aleksic}. Note also that $p=0.5$ superdiscs \textit{perfectly} cover (tile) the plane.

\begin{figure}

\begin{minipage}{4cm}
\includegraphics[width=4cm,height=4cm]{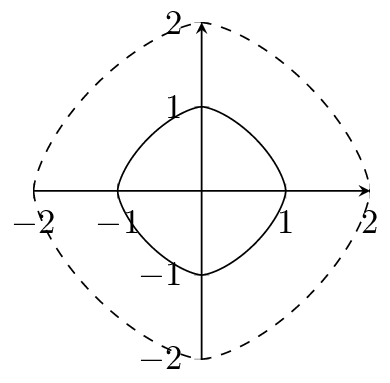}
\subcaption{}
\end{minipage}
\begin{minipage}{4cm}
\includegraphics[width=4cm,height=4cm]{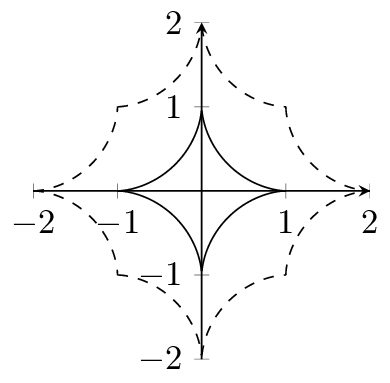}
\subcaption{}
\end{minipage}

\begin{minipage}{4cm}
\includegraphics[width=4cm,height=4cm]{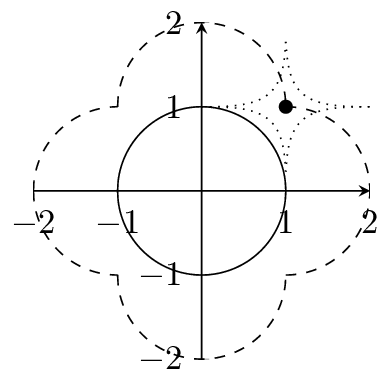}
\subcaption{}
\end{minipage}
\begin{minipage}{4cm}
\includegraphics[width=4cm,height=4cm]{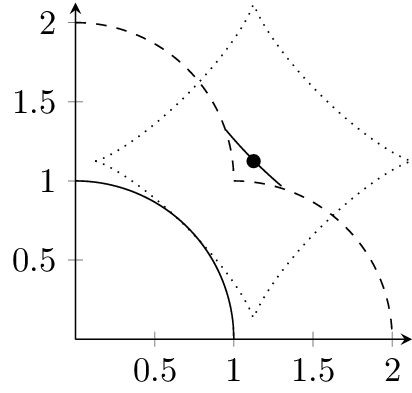}
\subcaption{}
\end{minipage}

\caption{Exclusion regions (dotted lines) for (a) convex objects, (b)concave objects, (c) mixture of concave and convex objects with $p<p_{c}$, and (d) ditto with $p>p_{c}$ (see text).}
\label{excreg}
\end{figure}

\begin{figure}
\includegraphics[width=8cm]{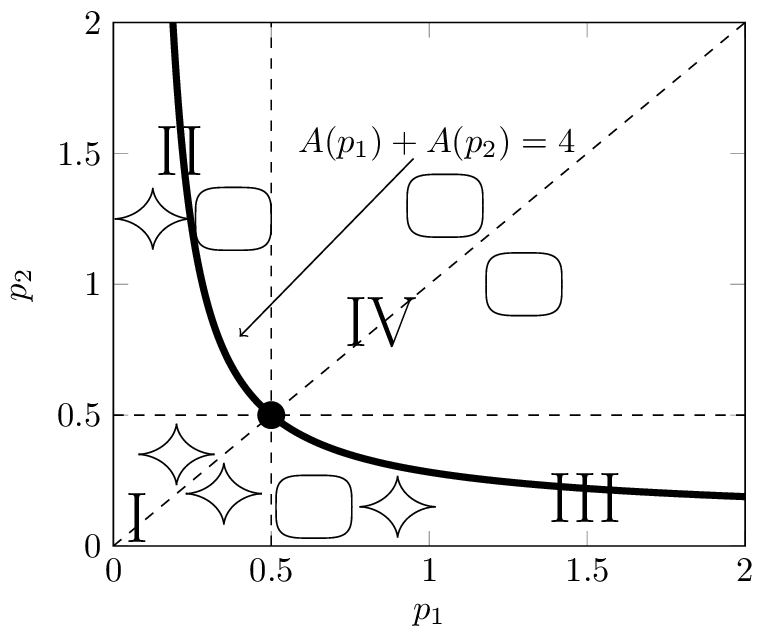}

\caption{Phase diagram of shape combinations for binary mixture $p_{1}$ and $p_{2}$. Shapes are represented symbolically by concave and convex symbols. Full black line is the locus of points where two shapes fit together perfectly to completely tile the plane.}
\end{figure}

In the case of binary mixtures, when two species of superdisks characterized by different deformation parameters $p_{1}$ and $p_{2}$ are deposited, the construction of the corresponding ER is as follows. To gain orientation, it is useful, first, to consider a $(p_{1},p_{2})$ plane of possible shape combinations. When both $p_{1}<0.5$ and $p_{2}<0.5$, the mixture is of two concave superdisks, denoted as Region I in $(p_{1},p_{2})$ plane and shown in Fig. 3. Similarly, Regions II and III  are regions of concave/convex shape combinations (i.e., $p_{1}<0.5, p_{2}>0.5$, and vice versa), while Region IV corresponds to convex/convex combination ($p_{1}>0.5, p_{2}>0.5$) as shown schematically in Fig. 3 (see figure).
In Region I, the concave construction (as described above for the pure case) must be used, but this time using the superdisk with \textit{larger} of the two parameters $p_{1}$ and $p_{2}$ in the mixture. Thus, the corresponding ER has the shape as illustrated in Fig. 2(a), its surface area is $\mathcal {A}_{cv}=4+2A(p)$ where $A(p)$ is given by eq. (1), with $p=max(p_{1},p_{2})$.
\begin{figure}
\includegraphics[width=8cm]{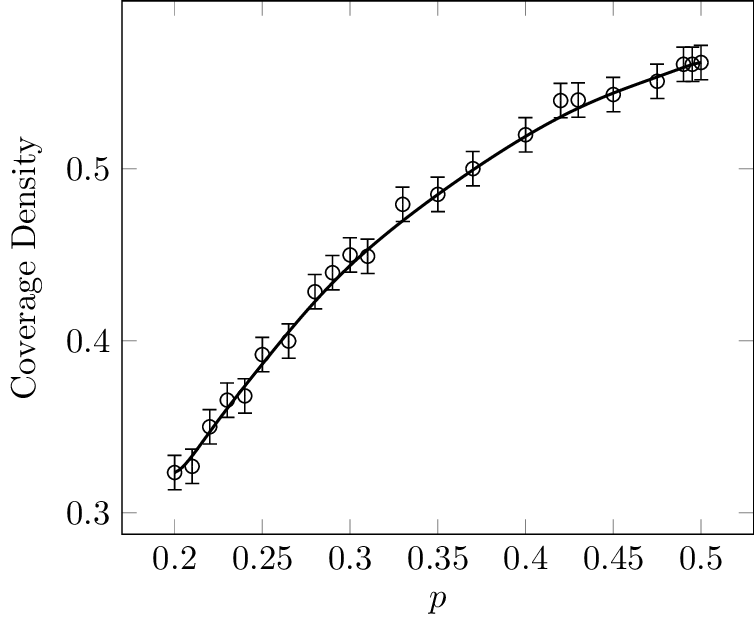}
\caption{Coverage density as a function of (smaller) deformation parameter $p$ along the locus of nonanalyticities.}
\end{figure}

In Region II (or, equivalently, Region III) of concave/convex mixtures, somewhat different ER construction must be used. For illustration, and withous loss of generality, consider the mixtures of circle ($p_{2}=1$, a convex object) with a concave superdisks ($p_{1}<0.5$). As long as $p_{1}$ is less than certain value, $p_{c}$ (see below), the appropriate ER is obtained by concave construction carried out with convex object (circle in this example). This is illustrated in Fig. 2(c), where ER consists of four semicircles obtained by unit translations of the original circle along $\pm x, \pm y$ axes. (The area of this ER is simply $4+2\pi$). In other words, concave object with, say, $p_{1}=0.2$ is "small enough" for its center to be placed at the four nearest points to the circle - points $(1,1), (1,-1),(-1,1)$, and $(-1,-1)$ - without overlap. This is shown in Fig. 2(c) for the case $p_{1}=0.2$. However, when $p_{1}>p_{c}$, say $p_{1}=0.4$, as illustrated in Fig. 2(d), this is not the case ( for better view, this illustration only shows the first quadrant, as the others are the same by symmetry). The corresponding ER must be modified near the point $(1,1)$ to accomodate for the extra bulge produced by the increased value of $p_{1}$ from 0.2 to 0.4. The exact form of this modification is geometricaly obtained by placing the center of $p_{1}$ object at the point $\left(\frac{1}{\sqrt2}, \frac{1}{\sqrt2}\right)$, or, in general case $\left(2^{-\frac{1}{2p}}, 2^{-\frac{1}{2p}}\right)$, and keeping the remaining segment between the intersections of its boundary with semi-circles of the ER. The quarter of the contour of the new ER is illustrated in Fig.2(d). The new ER obtained in this way has the area $4+2\pi+4\alpha(p_{1})$ where $\alpha(p_{1})$ is the area of the added triangular region, as seen in Fig. 2(d). The exact expression for this area can be derived, but the resulting equation is too long and cumbersome to be reproduced here (this and other derivations and numerical details will be published separately).

Clearly, there exists a value of $p_{1}=p_{c}$, when concave superdisk $p_{1}$ snugly fits with the circle ($p_{2}=1$), to fully cover the plane. In this example, the value is $p_{c}=-\frac{1}{2}\frac{\ln(2)}{\ln\left(1-\frac{\sqrt2}{2}\right)}=0.28223819..$. Also, at $p_{1}=p_{c}$, the ER construction must be changed, and, even though the area of ER is continuous function, it is nonanalytic because its left and right derivatives at $p_{c}$ are different. This is analogous to the result obtained for the "pure" case.  More generally, for arbitrary convex/concave mixture of superdisks, the pairs of $p_{1}$ and $p_{2}$ can be found for which the corresponding shapes snugly fit and ideally cover the plane. The exact expression for the locus of such pairs in $(p_{1},p_{2})$ plane can be derived, for example, by invoking Minkowski's convex body theorem from the geometry of numbers \cite{Minkowski}, and is given by $A(p_{1})+A(p_{2})=4$. Alternatively, one can derive equivalent simpler expression $2^{\frac{4p_{1}-1}{2p_{1}}}+2^{\frac{4p_{2}-1}{2p_{2}}}=4$ with the same solutions. This locus is marked by the full line in Fig. 3. Along that locus, the maxumum packing density equals 1.

As $p_{1}$ increases beyond $p_{c}$, the construction remains the same until $p_{1}=0.5$. At that point, $p_{1}$-superdisk becomes convex, and  we are in Region IV of shape combinations (the mixture of two convex shapes). The ER in this region is easily obtained in parametric form. Specifically, for $p_{1}>0.5$, $p_{2}>0.5$, and $p_{1}<p_{2}$, the boundary of the corresponding ER, in the first quadrant, is given by  $x(t)={(1+t^{2p_{1}})^{-\frac{1}{2p_{1}}}}+(1+t^{\frac{2p_{2}(2p_{1}-1)}{2p_{2}-1}})^{-\frac{1}{2p_{2}}}$, $y(t)={t\times(1+t^{2p_{1}})^{-\frac{1}{2p_{1}}}}+t^{\frac{2p_{1}-1}{2p_{2}-1}}\times(1+t^{\frac{2p_{2}(2p_{1}-1)}{2p_{2}-1}})^{-\frac{1}{2p_{2}}}$, and $t\in (0,\infty)$, with appropriately changed signs for $x(t)$ and $y(t)$ in the remaining quadrants. When $p_{1}>p_{2}$, the same expression applies with $p_{1}$ and $p_{2}$ interchanged. When $p_{1}=p_{2}=p$ (the pure case), upon eliminating parameter $t$, we recover $x^{2p}+y^{2p}=2^{2p}$, as expected.

In order to study the effect of mixing on coverage, we performed Monte Carlo simulations of deposition of mixture of superdisks that snugly fit, i.e., the pairs of shapes that lie on the locus of singularities. Results are shown in Fig. 4. where we plot coverage density vs. the value of smaller $p$ in the mixture. The results show that the mixture of ideally fitting superdisks covers the plane less efficiently than the pure case of $p=0.5$ objects. This intriguing result deserves further study, beyond the scope of this paper.
   
In conclusion, we have analyzed deposition of binary mixtures of variously shaped superdisks on planar surface and derived the exact form of phase boundary in the space of shape combinations. Numerical simulation of deposition of such objects indicates that the best values of jamming coverages are obtained when the objects are similar in shape. 
\begin{acknowledgments}
This publication was made possible by NPRP grants \#09 - 462 - 1 - 074 and \#5 - 674 - 1 -114 from the Qatar National Research Fund (a member of Qatar Foundation). The statements made herein are solely the responsibility of the authors. Work at the Institute of Physics Belgrade is supported by the Ministry
of Science of the Republic of Serbia under the projects OI 171006 and ON 171017.
\end{acknowledgments}

\end{document}